\documentclass[aps,twocolumn,prl,floatfix,showpacs,citeautoscript,superscriptaddress]{revtex4-1}

\usepackage[none]{hyphenat}
\usepackage{amssymb}
\usepackage{amsmath}
\usepackage{graphicx}
\usepackage{bm}
\usepackage{times}
\usepackage{color,soul}
\usepackage[dvipsnames,svgnames,table]{xcolor}
\usepackage{hyperref}
\usepackage{fancyhdr}
\usepackage{enumitem}
\usepackage[english]{babel}
\usepackage[caption=false]{subfig}
\usepackage{braket}
\usepackage{scalerel}

\usepackage{mathrsfs}
\usepackage{MnSymbol}
\usepackage{lipsum}

\hypersetup{
pdfstartview={FitH},% fits the width of the page to the window
colorlinks=true,    % false: boxed links; true: colored links
linkcolor=NavyBlue, % color of internal links
citecolor=Blue,   % color of links to bibliography
filecolor=NavyBlue, % color of file links
urlcolor=NavyBlue   % color of external links
}

\begin{document}

\title{Emergent parametric resonances and time-crystal phases in driven BCS systems
}

\author{H. P. {Ojeda Collado}}
\email{hector.pablo.ojedacollado@roma1.infn.it}
\affiliation{ISC-CNR and Department of Physics, Sapienza University of Rome, Piazzale Aldo Moro 2, I-00185, Rome, Italy}
\affiliation{Centro At{\'{o}}mico Bariloche and Instituto Balseiro,
Comisi\'on Nacional de Energ\'{\i}a At\'omica (CNEA)--Universidad Nacional de Cuyo (UNCUYO), 8400 Bariloche, Argentina}
\affiliation{Instituto de Nanociencia y Nanotecnolog\'{i}a (INN), Consejo Nacional de Investigaciones Cient\'{\i}ficas y T\'ecnicas (CONICET)--CNEA, 8400 Bariloche, Argentina}
\author{Gonzalo Usaj}
\affiliation{Centro At{\'{o}}mico Bariloche and Instituto Balseiro,
Comisi\'on Nacional de Energ\'{\i}a At\'omica (CNEA)--Universidad Nacional de Cuyo (UNCUYO), 8400 Bariloche, Argentina}
\affiliation{Instituto de Nanociencia y Nanotecnolog\'{i}a (INN), Consejo Nacional de Investigaciones Cient\'{\i}ficas y T\'ecnicas (CONICET)--CNEA, 8400 Bariloche, Argentina}
\author{C. A. Balseiro}
%\email[Corresponding author: ]{balseiro@cab.cnea.gov.ar}
\affiliation{Centro At{\'{o}}mico Bariloche and Instituto Balseiro,
Comisi\'on Nacional de Energ\'{\i}a At\'omica (CNEA)--Universidad Nacional de Cuyo (UNCUYO), 8400 Bariloche, Argentina}
\affiliation{Instituto de Nanociencia y Nanotecnolog\'{i}a (INN), Consejo Nacional de Investigaciones Cient\'{\i}ficas y T\'ecnicas (CONICET)--CNEA, 8400 Bariloche, Argentina}
\author{Dami{\'{a}}n H. Zanette}
\affiliation{Centro At{\'{o}}mico Bariloche and Instituto Balseiro,
Comisi\'on Nacional de Energ\'{\i}a At\'omica (CNEA)--Universidad Nacional de Cuyo (UNCUYO), 8400 Bariloche, Argentina}
\affiliation{Consejo Nacional de Investigaciones Cient\'{\i}ficas y T\'ecnicas (CONICET), Argentina}
\author{Jos\'{e} Lorenzana}
\email{jose.lorenzana@cnr.it}
\affiliation{ISC-CNR and Department of Physics, Sapienza University of Rome, Piazzale Aldo Moro 2, I-00185, Rome, Italy}

\begin{abstract}
  We study the out-of-equilibrium dynamics of a Bardeen-Cooper-Schrieffer condensate subject to a periodic drive. We demonstrate that the combined effect of drive and interactions results in emerging parametric resonances, analogous to a vertically driving pendulum. In particular, Arnold tongues appear when the driving frequency matches $2\Delta_0/n$, with $n$ a natural number, and $\Delta_0$ the equilibrium gap parameter. Inside the Arnold tongues we find a commensurate time-crystal condensate which retains the $U(1)$ symmetry breaking of the parent superfluid/superconducting phase and shows an additional time-translational symmetry breaking. Outside these tongues, the synchronized collective Higgs mode found in quench protocols is stabilized without the need of a strong perturbation. Our results are directly relevant to cold-atom and condensed-matter systems and do not require very long energy relaxation times to be observed. 
\end{abstract}

\date{\today}
\maketitle

%%%%%%%%%%%%%%%%%%%%%%%%%%%%%%%%%%%%%%%
%\section{Introduction}
%%%%%%%%%%%%%%%%%%%%%%%%%%%%%%%%%%%%%%%
 
Periodic driving of a many-body system allows the manipulation of the equilibrium phase diagram through phenomena such as dynamic localization~\cite{Zenesini2009,Struck2011,Struck2013},
and the creation of new out-of-equilibrium states of matter as time-crystals~\cite{Zhang2017b,Choi2017,Autti2018,Kyprianidis2021} which exhibits time-translational symmetry breaking (TTSB).  Related effects have been found in solids, often in connection with an enhanced stability of broken symmetry phases~\cite{Fausti2011,Mitrano2016,Beck2013,
Nova2019,Kogar2020,Basov2017}. %  or hidden phases~\cite{Stojchevska2014}. 
These phenomena can be described using mathematical techniques developed by Floquet in the nineteenth century~\cite{floquet1883} 
and referred to as Floquet engineering~\cite{Oka2019}.

An interesting Floquet engineering technique is to exploit parametric resonances and the associated parametric amplification allowing, for example, to overcome intrinsic losses~\cite{Ma2019}. Quite generally, parametric resonances require a non-linear medium whose intrinsic parameters can be modified periodically by a drive. This can be achieved in metamaterials as an array of Josephson junctions~\cite{Castellanos-Beltran2008,Bergeal2010,Macklin2015} or a structured superconducting wave-guide~\cite{HoEom2012}. Parametric resonances can also be generated in a single material as, for example, a layered superconductor with intrinsic Josephson coupling between planes~\cite{Rajasekaran2016,Homann2020,VonHoegen2021}, a superconductor with surface allowed non-linear coupling with the electromagnetic field~\cite{Buzzi2021}, a charge-density-wave~\cite{Liu2013} with non-linear coupling between amplitude and phase modes, or a semiconductor with non-linear coupling between electric field and phonons~\cite{Cartella2018}.

In this Letter we show that parametric resonaces emerge naturally {\em in  any system} whose dynamics is described by the Bardeen-Cooper-Schrieffer (BCS) Hamiltonian at times short respect to the energy relaxation time (the so called pre-thermal regime).
% The dynamical phase diagram for subgap excitation shows a sequence of collective dynamical phase transitions (DPT) emulating the phase diagram of a single parametric oscillator.
Furthermore, we show that parametric resonances stabilize a discrete time-crystal phase~\cite{Wilczek2012,Yao2017,Russomanno2017,Huang2018,Khasseh2019,Pizzi2020,Pizzi2021,Kozin2019,Yang2021} which retains the $U(1)$ gauge symmetry breaking of the equilibrium BCS condensate.

 We consider a weak-coupling fermionic condensate with s-wave pairing described by the BCS model and subject to a periodic drive that couples with the order parameter. 
The system is treated in the Anderson pseudospin formulation~\cite{Anderson1958}. The Hamiltonian takes the form
%%%%%%%%%%%%
\begin{equation}\label{eq:ham}
\hat{H}_{\mathrm{BCS}}=-2\sum_{\bm{k}} \xi_{\bm{k}}\hat{S}_{\bm{k}}^{z}-\lambda(t)\sum_{\bm{k},\bm{k}'}\hat{S}_{\bm{k}}^{+}\hat{S}_{\bm{k}'}^{-}\,.
\end{equation} 
%%%%%%%%%%%%
Here, $\xi_{\bm{k}}=\varepsilon_{\bm{k}}-\mu$ measures the energy from the Fermi level $\mu$ and the pairing interaction 
is taken periodic in time with driving strength $\alpha$; $\lambda(t)=\lambda_0\,[1+\alpha\, \sin(\omega_d t)]$. The $\frac{1}{2}$-pseudospin operators are given by
%%%%%%%%%%%%%%%%%%%
%\begin{eqnarray}
%\nonumber
$\hat{S}_{\bm{k}}^{x}=\frac{1}{2}\left(\hat{c}_{\bm{k}\uparrow}^{\dagger}\hat{c}_{-\bm{k}\downarrow}^{\dagger}+\hat{c}_{-\bm{k}\downarrow}^{}\hat{c}_{\bm{k}\uparrow}^{}\right)$\,,
$\hat{S}_{\bm{k}}^{y}=\frac{1}{2i}\left(\hat{c}_{\bm{k}\uparrow}^{\dagger}\hat{c}_{-\bm{k}\downarrow}^{\dagger}-\hat{c}_{-\bm{k}\downarrow}^{}\hat{c}_{\bm{k}\uparrow}^{}\right)$\,, %\label{eq:sz}
$ 
\hat{S}_{\bm{k}}^{z}=\frac{1}{2}\left(1-\hat{c}_{\bm{k}\uparrow}^{\dagger}\hat{c}_{\bm{k}\uparrow}^{}-\hat{c}_{-\bm{k}\downarrow}^{\dagger}\hat{c}_{-\bm{k}\downarrow}^{}\right)$ %\nonumber
%\,,
%\end{eqnarray}
%%%%%%%%%%%%%%%%%
and $\hat{c}^\dagger_{\bm{ k}\sigma}$ ($\hat{c}_{\bm{k}\sigma}$) is the usual creation (annihilation) operator for fermions with momentum $\bm{k}$ and spin $\sigma$. The operator $\hat{S}_{\bm{k}}^{\pm}\equiv \hat{S}_{\bm{k}}^{x}\pm i\hat{S}_{\bm{k}}^{y}$ creates or annihilates a Cooper pair $(\bm{k},-\bm{k})$. Due to the all-to-all interaction in Eq.~(\ref{eq:ham}), a mean-field treatment becomes exact in the thermodynamic limit and the dynamics can be obtained solving for each pseudospin in a self-consistent field. 
The BCS mean-field Hamiltonian can be written as,
%%%%%%%%%%%%%%%%%
%\begin{equation}\label{eq:hamMF}
$\hat{H}_{\mathrm{MF}}=-\sum_{\bm{k}}\hat{\bm{S}}_{\bm{k}}\cdot\bm{b}_{\bm{k}}$, 
%\end{equation}
%%%%%%%%%%%%%%%%%
where, $\bm{b}_{\bm{k}}\left(t\right)=(2\Delta\left(t\right),0,2\xi_{\bm{k}})$ represents an effective magnetic field vector for the $\frac{1}{2}$-pseudospin operator $\hat{\bm{S}}_{\bm{k}}=(\hat{S}_{\bm{k}}^{x},\hat{S}_{\bm{k}}^{y},\hat{S}_{\bm{k}}^{z})$. Here, without loss of generality, we have assumed a real equilibrium BCS order parameter ($\Delta_0$), a condition that remains valid over time due to electron-hole symmetry. The instantaneous BCS order parameter is given by 
%%%%%%%%%%%%%%%%%%%%%%%
\begin{equation}
\label{eq:deltat}
\Delta(t)=\lambda(t)\sum_{\bm{k}} S_{\bm{k}}^{x}\,,
\end{equation}
%%%%%%%%%%%%%%%%%%%%%%%%%
where symbols 
%$S_{\bm{k}}^{x}$,
without hat denote the expectation value of operators
%$\hat{S}_{\bm{k}}^{x}$
in the time-dependent BCS state. % (this notation is used hereafter). 

At equilibrium, in the absence of periodic perturbations, the $\frac{1}{2}$-pseudospins align in the direction of their local fields $\bm{b}_{\bm{k}}^0=(2\Delta_{0},0,2\xi_{\bm{k}})$ in order to minimize the system's energy. %[described by  Eq.~(\ref{eq:hamMF})].
% This corresponds to the zero-temperature paired ground state in which the pseudospin texture (the expectation value of pseudospin operators as a function of momentum $\bm{k}$) is given by
% %%%%%%%%%%%%%%%%%%
% \begin{equation}
% \label{eq:sequilib} 
% S_{\bm{k}}^{x,0} =\frac{\Delta_{0}}{2\sqrt{\xi_{\bm{k}}^{2}+\Delta_{0}^{2}}},\: S_{\bm{k}}^{y,0} =0, \: S_{\bm{k}}^{z,0} =\frac{\xi_{\bm{k}}}{2\sqrt{\xi_{\bm{k}}^{2}+\Delta_{0}^{2}}}.
% \end{equation}
% %%%%%%%%%%%%%%%%%
This is used as initial condition and once the pairing interaction is modulated, the %expectation values of the
pseudospins evolve in time obeying the equation of motion
%%%%%%%%%%%%%%%%%%%%
%\begin{equation}
%\label{eq:eom}
$\frac{d\bm{S}_{\bm{k}}}{dt}=-\bm{b}_{\bm{k}}\left(t\right)\times\bm{S}_{\bm{k}}$
%\end{equation}
%%%%%%%%%%%%%%%%
%where we set
($\hbar\equiv1$).
In contrast with Refs.~\cite{OjedaCollado2018,OjedaCollado2020,Homann2020}, we focus on subgap frequencies ($\omega_d\le2\Delta_0$) and weak driving amplitude 
($\alpha<0.2$). We consider $N=10^4$ pseudospins equally spaced in energy  $\xi_{\bf k}$, within an energy range of $W=40\Delta_0$ around $\mu$~\cite{Note1}.
%with a constant density of states~\cite{Note1}. %$\nu$.

\begin{figure}[tb]
\includegraphics[width=0.4\textwidth]{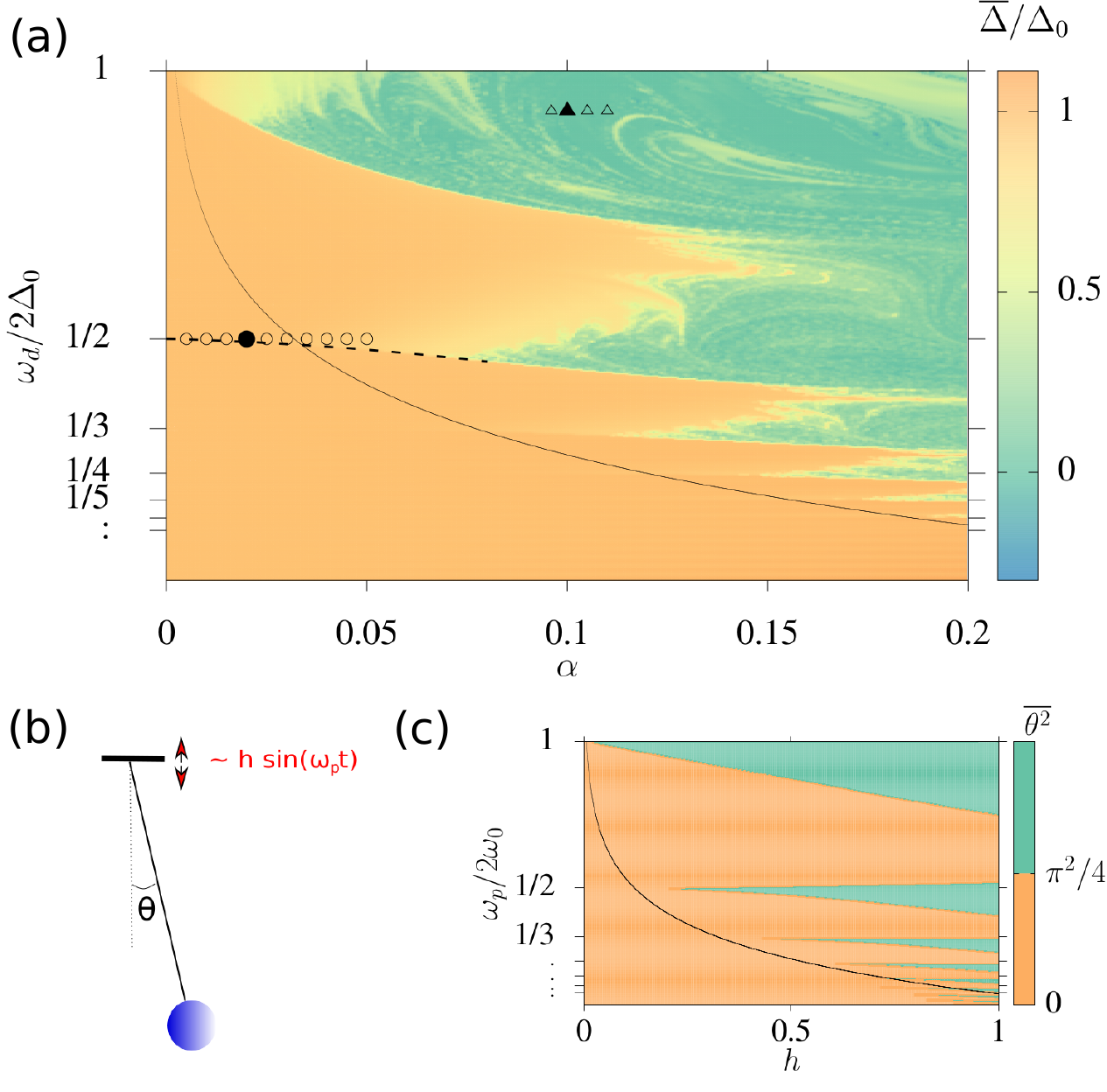}
\caption{(Color online)
   (a) Dynamical phase diagram obtained as  false colour plot of $\overline{\Delta}$ as a function of $\alpha$ and $\omega_d$. The order parameter was averaged in the interval $t\Delta_0 \in [0, 200]$.  Parametric resonances occur at  $2\Delta_0/n$ with $n$ a natural number similar to the vertically driven pendulum (b). The phase diagram of the latter is shown in (c) by plotting the mean-square amplitude of oscillation as a function of pump frequency $\omega_p$ and amplitude $h$ (in the harmonic approximation).  
   The black curves in (a) and (c) show a power law delimiting the Arnold tongues in the presence of damping. %(see Appendix~\ref{app:ap1}).
   Small discrepancies in panel (c) are finite time-window effects. In the case of the BCS system, we found that  $\alpha=0.45\eta^ {\omega_d/2\Delta_0}$ with $\eta=0.005$
approximately describes the numerical results.
The dashed black line in (a) indicates a weak-first order DPT. The open symbols indicate the parameters for which the rigidity of the Higgs mode (circles) and time-crystal (triangles) has been checked~\cite{Note1}. 
%Fig.~\ref{fig:fig4} and Appendix~\ref{app:ap5}.)
 } 
\label{fig:fig1}
\end{figure}

%\subsection{Phase diagram}\label{sec:phase-diagram}
 To characterize the dynamical phase transitions (DPT), we use $\overline{\Delta}$ as dynamical order parameter, defined as  
 the average of the order parameter $\Delta(t)$ over a large time window in the stationary regime. Figure~\ref{fig:fig1}(a) shows a map of  $\overline{\Delta}$ as a function of the driving strength $\alpha$ and driving frequency $\omega_d$. There are two distinct main regions in the phase diagram, one in which $\overline{\Delta}\approx 0$ (green area), and another one in which the temporal average is close to the initial equilibrium value $\Delta_0$ (orange area). 
We see resonant behaviour (i.e. the average order parameter close to zero),  each time the drive frequency matches $\omega_d=2\Delta_0/n$, with $n$ a natural number, within a region that becomes larger as the driving strength $\alpha$ grows,  forming so-called Arnold tongues. Since these features emerge varying an internal parameter of the system, they may be called parametric resonances. Even more substantially, the phase diagram is remarkably similar to that
in Fig.~\ref{fig:fig1}(c) corresponding to the archetypal model of a  parametric oscillator, namely, a vertically forced pendulum  with a pump frequency $\omega_p$ as shown schematically in (b).
This analogy requires:  i) to identify the natural frequency as $\omega_0=2\Delta_0$, ii) to identify the pump frequency of the pendulum as $\omega_p=2\omega_d$, iii) to include a small damping constant $\eta$ in the pendulum~\footnote{See Supplemental Material}, and iv) to identify the regions with large deviation from equilibrium in the pendulum (resonances) with the regions of zero average gap. Thus, the usual parametric resonances at $\omega_p=2\omega_0/n$ correspond one-to-one to the resonances we observe in the BCS system. 

In the damped pendulum, the Arnold tongue starts at a sharp value of  the driving strength~\cite{Landau1976} satisfying a power law $h_c=\eta^{1/n}$. This is approximately verified for the BCS system (thin black line) but the behaviour is more complex. The upper boundary of the Arnold tongue has a fractal-like structure similar to the chaotic dynamics found in related systems~\cite{Lerose2018,Lerose2019}. In contrast,
%On the other hand,   
%analysing in detail one particular case  ($n=2$), we find that
the lower boundary does not finish at the tip of the Arnold tongue but continues until $\alpha=0$ as a weak first-order DPT where $\overline{\Delta}$ changes discontinuously (dashed line for $n=2$) producing a sharp edge.

%(see Appendix~\ref{app:ap3}).
 %( Fig.~\ref{fig:figs1}).

%%%%%%%%%%%%%%%%%%%%%%%%%%%%%%%%%%%%%%%%%%%%%%%%%%%%%
%%%%%%%%%%%%%%%%%%%%%%%%%%%%%%%%%%%%%%%%%%%%%%%%%%%%%%%%

\begin{figure}[tb]
\includegraphics[width=0.5\textwidth]{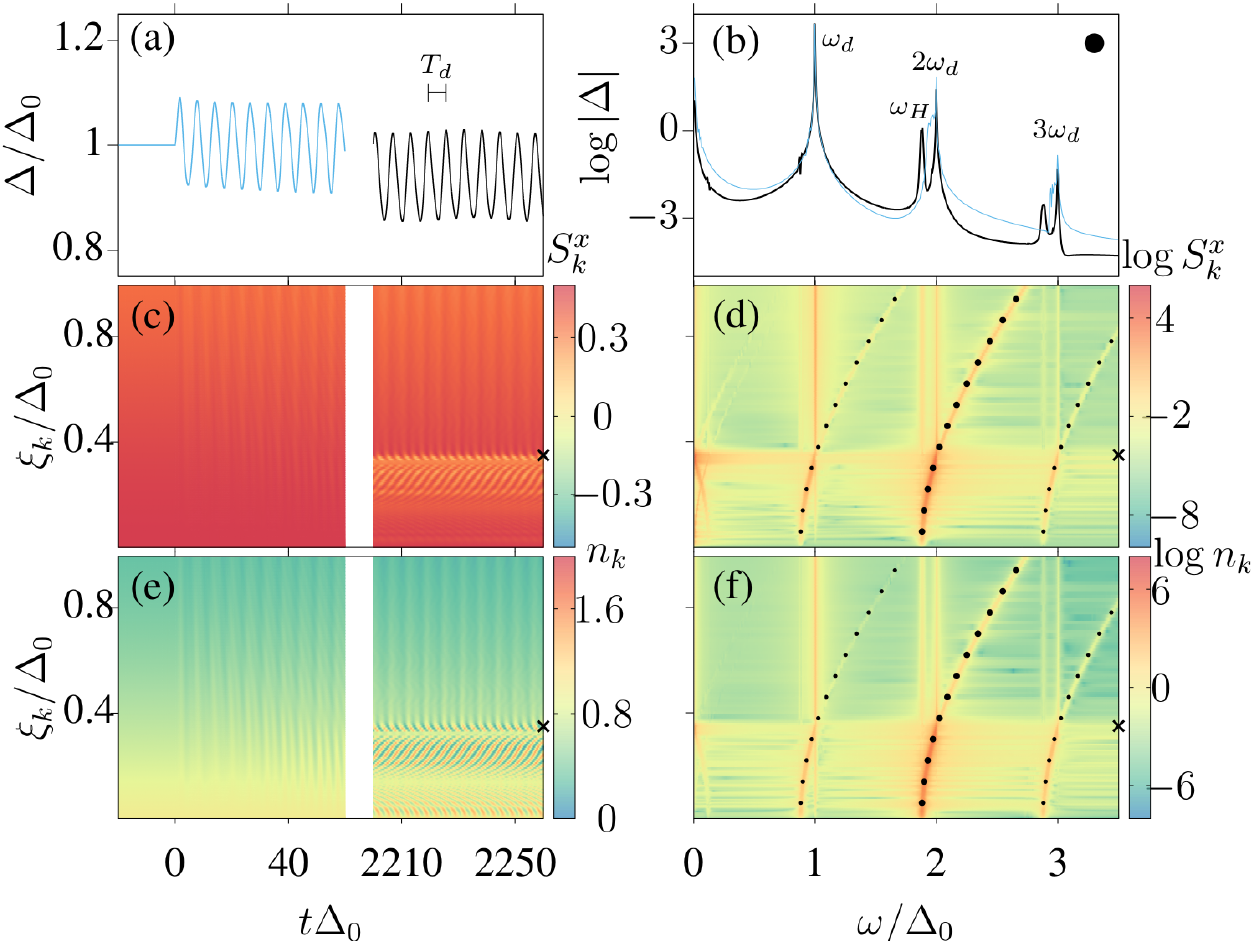}
\includegraphics[width=0.2\textwidth]{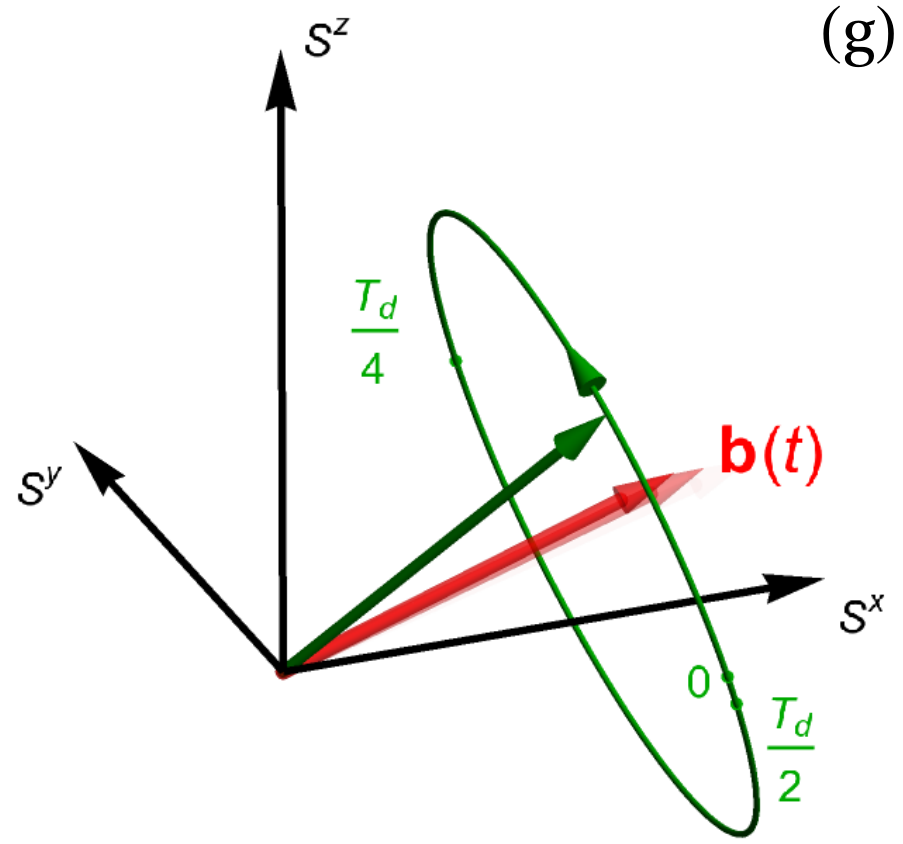}
\includegraphics[width=0.25\textwidth]{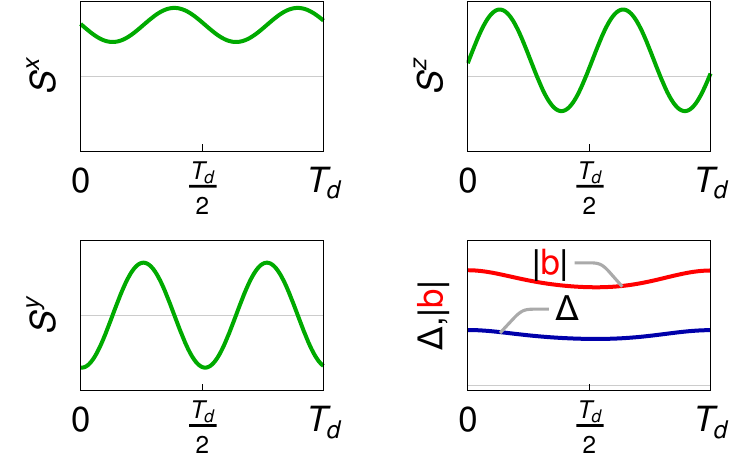}
\caption{(Color online) The dynamics outside the Arnold tongues 
  [corresponding to the filled circle in Fig.~\ref{fig:fig1}(a)] is characterized through $\Delta(t)$, $S_{\bm{k}}^{x}(t)$ and $n_{\bm{k}}(t)$. %(here we used $t\Delta_0 \in [1500, 2500]$ for high frequency resolution).
  Panels (a),(c) and (e) show the transient dynamics (left) and the steady state dynamics (right).
  Panels (b), (d) and (f) shows the log of the FT.
  Black dots in panels (d) and (f) mark 
%  correspond to %the effective Larmor frequency
$\Omega_L$
%and satellites at
and $\Omega_L\pm \omega_d$. The cross at the right of panels (c)-(f) indicate the boundary of the $2\omega_d$-resonance regime.  
Panel (g) illustrates schematically the pseudospin precessions around the pseudomagnetic field $\bm{b}(t)$  which has a time dependence through $\Delta (t)$ ($\bm{k}$ label dropped for clarity). The chosen pseudospin with $\xi_{\bm{k}}/\Delta_0=0.3$ is in the 
$2\omega_d$-resonance regime,
so its Larmor frequency is nearly twice the periodicity of $\Delta (t)$ and the modulus of $\bm{b}(t)$ as shown on the right.} 
\label{fig:fig2}
\end{figure}

To characterize the different dynamics outside and inside the Arnold tongues,
in the following we analyse in more detail two representative examples.
Figure \ref{fig:fig2} shows the dynamics away from the Arnold tongues [full circle in Fig.~\ref{fig:fig1}(a)]. After some transient oscillations, the superconducting order parameter decreases in average and oscillates around a new value,  $\overline{\Delta}<\Delta_0$ [Fig.~\ref{fig:fig2}(a)]. We have also plotted the $x$-component of the pseudospin texture over time $S_{\bm{k}}^{x}(t)$ [Fig.~\ref{fig:fig2}(c)], which is associated to the superconducting response through Eq.~(\ref{eq:deltat}), as well as the time-dependent quasiparticle distribution $n_{\bm{k}}(t)=1-2 S_{\bm{k}}^{z}(t)$ [Fig.~\ref{fig:fig2}(e)].  The Fourier transforms (FT) in panels (d) and (f) show that the same set of frequencies appear in the dynamics of $n_{\bm{k}}$ and $S_{\bm{k}}^{x}$. As illustrated in Fig.~\ref{fig:fig2}(g), this is a simple consequence of the fact that pseudospins (green) make a tilted precession around the self-consistent pseudomagnetic field (red). As could be expected, the pseudospin Larmor frequency $\Omega_L$ is determined by the 
average gap (as opposed to the equilibrium gap). Indeed, the large dots in panels (d) and (f) mark $\omega=\Omega_{L}(\xi_{\bm k})\equiv 2\sqrt{\xi_{\bm{k}}^2+\overline{\Delta}^2}$.
Also, Floquet sidebands appear at $\Omega_{L}\pm \omega_d$ (small dots).

The cross at the right of panels (c)-(f) indicate the value  $\xi_{\bm{k}}^*$
satisfying $2\omega_d=\Omega_{L}(\xi_{\bm{k}}^*)$. Quasiparticles with $\xi_{\bm{k}}\lesssim \xi_{\bm{k}}^*$ are driven strongly out of equilibrium, creating a sharp separation in $\xi_{\bm{k}}$ among quasiparticles that respond strongly and weakly to the drive. 
%(see Supplementary Video~1 for more  details on the dynamics).
We will refer to this as $2\omega_d$-resonant behavior to be distinguish from parametric resonances. 

%Notwithstanding the strongest response below  $\xi_{\bm{k}}^*$ is dominated by the $2\omega_d$-resonant behaviour, the total FT shown in panel (b) is dominated by the response at the driving frequency $\omega_d$, which is weaker but effective in a wider range of the spectrum.

%On the other hand, we anticipate that the $2\omega_d$-resonant response plays a dominant role in the DPTs. 

%%%%%%%%%%%%%%%%%%%%%%%%%%%%%%%%%%%%%%%%%%%%%%%%%%%%%
%%%%%%%%%%%%%%%%%%%%%%%%%%%%%%%%%%%%%%%%%%%%%%%%%%%%%%%%

The  peak at $\omega_d$ and higher harmonics, shown in panel (b), can be explained from linear response theory and weak nonlinearities. In contrast, an unexpected oscillation occurs with a frequency $\omega_H$ which is not commensurate with the driving frequency but instead satisfies
%it is determined by the order parameter, 
$\omega_H=2\overline{\Delta}$. Thus, it is an internal mode of the many-body system that spontaneously emerges in the dynamics (Higgs mode). 
Being the time analogue of an incommensurate charge-density wave in a solid, these states are dubbed time quasicrystals in other contexts where they have been identified both experimentally~\cite{Autti2018} and theoretically~\cite{Giergiel2019,Volovik2013,Homann2020}. Here, we find that its frequency is robust to changes in the drive~\cite{Note1} when measured in units of $2\overline{\Delta}$ %(see Appendix~\ref{app:ap4} for details)
which is a general requirement defining time-crystal behaviour. On the other hand, since time-crystal is often associated to a subharmonic response, we use the conventional denomination of ``synchronized Higgs mode'' keeping in mind that it shares many characteristics of time-crystal behavior.

The vertical features in panels Fig.~\ref{fig:fig2}(d) and (f) at $\omega=\omega_H$ reveal that the origin of the synchronized Higgs mode is not a simple consequence of the Van Hove singularity of the BCS density of states, but of a synchronization between a group of pseudospins. 
A similar oscillation emerges spontaneously in BCS quench protocols in which the attractive interaction is  suddenly increased by a large amount~\cite{Barankov2004,Barankov2006a,Scaramazza2019,Gambassi2011}. There, also, the frequency of this Higgs mode is determined by the average gap~\cite{Seibold2020}. Here, the synchronized Higgs mode emerges with a continuous wave pump and without the need of large driving amplitudes, a protocol which is much  easier to implement experimentally~\cite{Behrle2018}.

Excitation of the Higgs mode by a periodic drive {\em above} the equilibrium gap 
was found in a layered Ginzburg-Landau model %including coupling to Josephson plasmons but
without quasiparticle excitations~\cite{Homann2020}. Our result applies to general BCS systems for driving frequency below the gap (where heating effects are expected to be minimized) and takes into account the full BCS dynamics including the effect of quasiparticle excitations.

%\subsection{Dynamics inside the Arnold tongues: commensurate time-crystal condensate}\label{sec:reson-dynam-comm}

\begin{figure}[tb]
\includegraphics[width=0.5\textwidth]{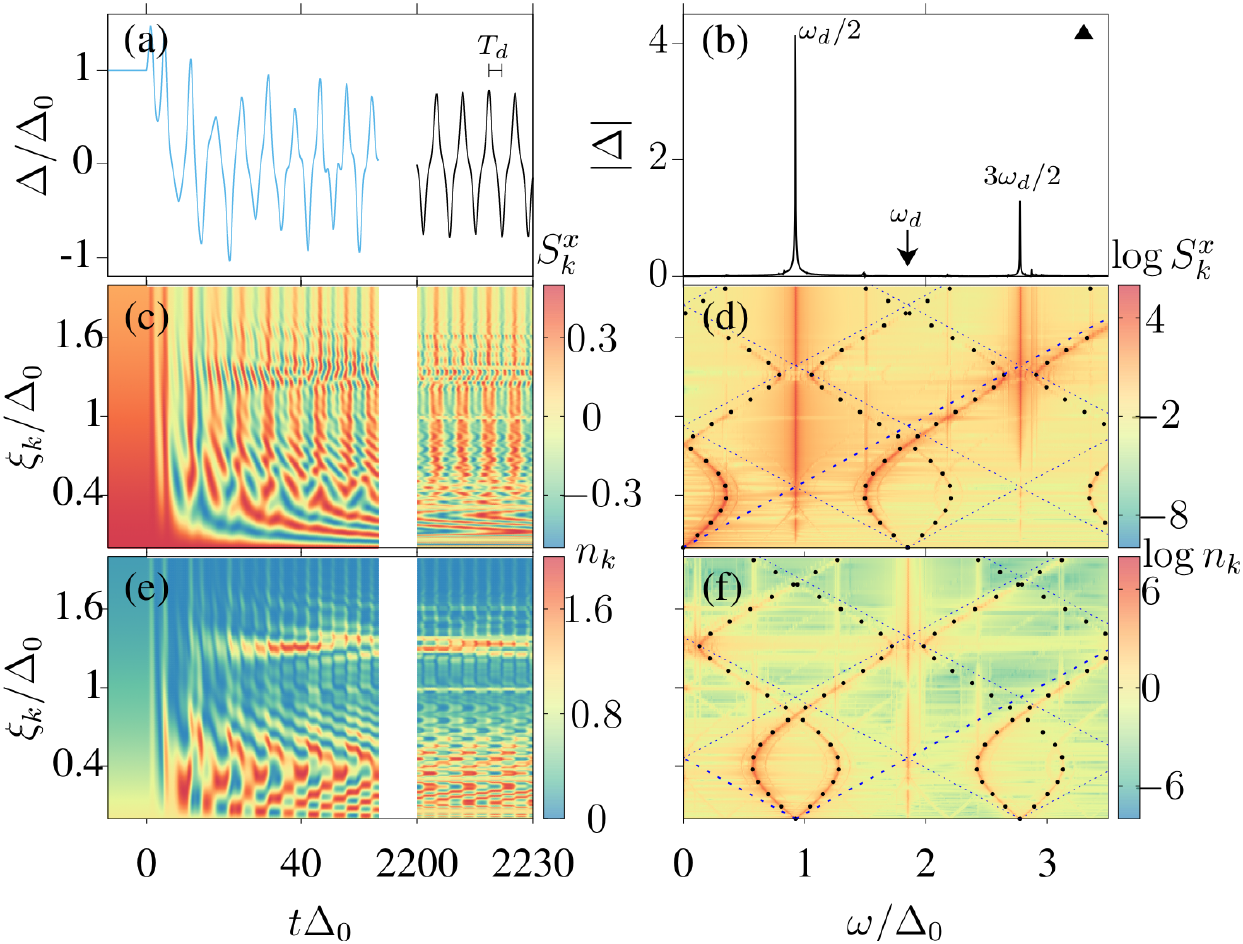}
\includegraphics[width=0.2\textwidth]{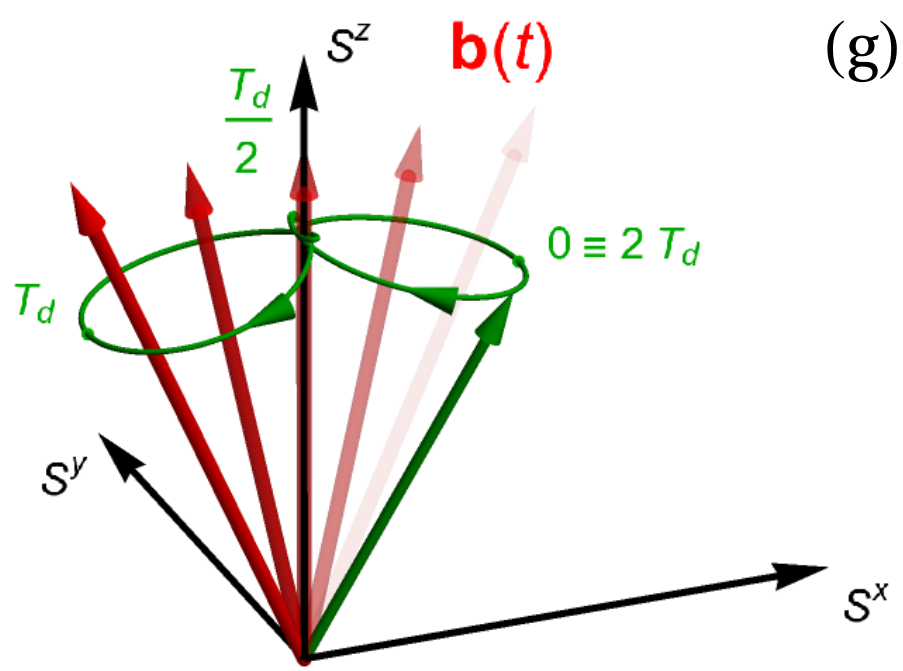}
\includegraphics[width=0.25\textwidth]{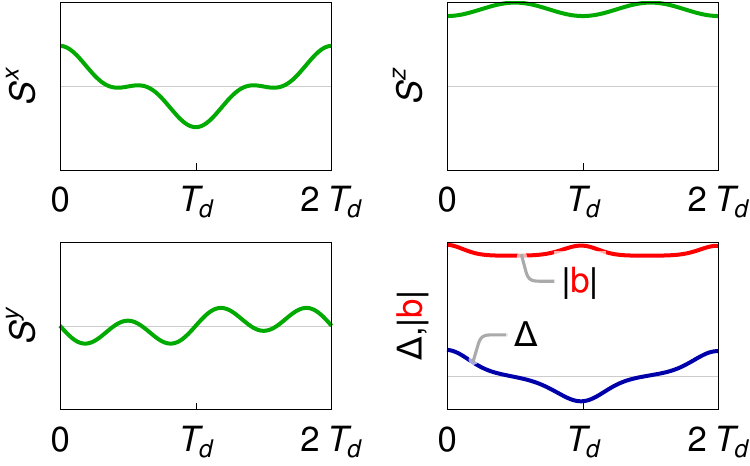}
\caption{(Color online) Same as Fig.~\ref{fig:fig2} but inside the Arnold tongue showing subharmonic dynamics [parameters corresponding to the filled triangle in Fig.~\ref{fig:fig1}(a)].
After a short transient, the average order parameter is driven to zero and oscillates with twice the periodicity of the drive [panels (a) and (g)]. Panel (g) schematizes the behaviour of a quasiparticle with $\xi_{\bm{k}}/\Delta_0=1.8$.
The pseudomagnetic field  (red) oscillates 
with a time-dependent length $|b|$ reminiscent of the vertically driven pendulum setting, i.e. the length is minimum each time the field is vertical ($t=T_d/2, 3T_d/2$) (red line in the lower right panel). The pseudospin (green) precesses and follows $\bm{b}(t)$, making an `$\infty$-shaped' loop. 
% (see also %Appendix~\ref{app:ap5} and Supplementary Video~2).
$S^x$ has $2T_d$ periodicity and self-consistently builds $\Delta(t)$ with the same periodicity. The dots in (d) are the result of a Floquet computation~\cite{Note1}. %(Appendix~\ref{app:ap6}).
The dashed blue lines are the bare dispersion $\omega=|2\xi_{\bm k}|$ (more intense) and associated Floquet side bands $|2\xi_{\bm k}\pm n\omega_d|$ with $n=1,2,3,...$
Notice that the Floquet spectrum appear shifted by $\omega_d/2$ in the charge response (f) in relation to panel (d). 
} 
\label{fig:fig3}
\end{figure}

We now switch to the typical dynamics inside the Arnold tongues (filled triangle in Fig.~\ref{fig:fig1}). Remarkably, we find that a new commensurate  time-crystal condensate phase emerges.
Indeed, as shown in Fig.~\ref{fig:fig3}, after a short transient, $\overline\Delta$ becomes zero [Fig.~\ref{fig:fig3}(a)]  
and the instantaneous order parameter oscillates with \textit{half} of the drive frequency [Fig.~\ref{fig:fig3}(b)] as found in other models showing discrete time-translational symmetry breaking~\cite{Sacha2015,Else2016,Chandran2016,Zhang2017b,Choi2017,Russomanno2017,Heugel2019,Yao2020,Natsheh2021,Natsheh2021b}.
%Notice that there is no response of the order parameter at the drive frequency, but only odd multiples of $\omega_d/2$ appear. This is due to an emerging symmetry fulfil by the steady state response upon time translation by one driving period, namely $\Delta(t)=-\Delta(t+T_d)$ with  $T_d=2\pi/\omega_d$.

%%%%%%%%%%%%%%%%%%%%%%%%%%%%%%%%%%%%%%%%%%%%%%%%%%%%%
%%%%%%%%%%%%%%%%%%%%%%%%%%%%%%%%%%%%%%%%%%%%%%%%%%%%%%%%

Figure~\ref{fig:fig3}(d) shows that this subharmonic response is shared by a wide range of pseudospins as witnessed by the vertical feature at $\omega_d/2$. This leads to a TTSB macroscopic response of the order parameter.  
We have checked that ``long-range'' order, in the sense of long autocorrelation time, holds at least for thousands of Floquet cycles in our numerical calculations. 

Figure~\ref{fig:fig3}(b) show that $\Delta\left(t\right)$ can be well approximated by only two Fourier components: $\Delta\left(t\right)=\Delta_{1}\cos\left(\omega_{d}t/2\right)+\Delta_{2}\cos\left(3\omega_{d}t/2\right)$ with $\Delta_{1}=0.58\Delta_{0}$ and $\Delta_{2}=0.156\Delta_{0}$. We can therefore use the  Bloch-Floquet theorem to analyze the corresponding pseudospin dynamics~\cite{Note1} which results in the dots shown in panels (d) and (f). These are given by twice the single-particle Floquet quasienergies %reported in Appendix~\ref{app:ap6}
and agree very well with the structures seen in the numerical simulations.

Near $\xi_{\mathbf{k}}/\Delta_0\approx 0.4$ a large gap of size $2\Delta_{1}$ appears in the spectrum of panel (d) as the remnant of the superconducting pairing. 
However, rather than being centered at $\omega=0$ [as for a conventional superconducting phase, c.f. Fig.~\ref{fig:fig2}(d)], 
it is centered at $\omega_{d}/2$.
Such finite frequency gap can be understood as the avoiding crossing among the bare dispersion  $\omega=|2\xi_{\mathbf{k}}|$  and the first Floquet replica $\omega=|\omega_d\pm 2\xi_{\mathbf{k}}|$ (dashed blue lines). Avoiding crossings involving higher Floquet replicas explain the smaller gaps of size $2\Delta_{2}$ around $\xi_k/\Delta_0\approx 1.3$.

%From a direct visualization of the dynamics, we find that quasiparticles with $\xi_{\mathbf{k}}$ matching these dynamical gaps, have strong pairing correlations (strong $x$ and $y$ components of the pseuodspins) but with different phases (see Supplementary Video~2). On the other hand, pseudospins remain active at arbitrary low frequencies, consistent with a gapless state.

Panel (g) schematizes the dynamics at high energy,
%(see also Supplementary Video~2),
where the subharmonic response is strong. Pseudospins precess and follow the time-dependent field that they contribute to create. Because the $z$-axis coincides with a symmetry axis of the dynamics, different frequencies appear in the pairing (d) and charge (f) fluctuations, in contrast to
% to the case analysed in
Fig.~\ref{fig:fig2} where such symmetry does not hold. Notice that charge fluctuations (f) respond at the drive frequency, in contrast to the subharmonic response of the pairing fluctuations (d). 

Due to the gauge invariance of the equations, multiplying the real $\Delta(t)$ by an arbitrary time-independent phase factor, $\Delta(t)e^{i\phi}$, yields another solution of the time-dependent BCS problem. In other words, the dynamical phase breaks both discrete time-translational symmetry and $U(1)$ symmetry. The latter symmetry breaking characterizes also an equilibrium BCS condensate thus the new state is dubbed a ``commensurate time-crystal condensate''. 

% Because the BCS dynamics is mean-field like, it can be mapped to the classical dynamics of a system of coupled non-linear oscillators, one for each pseudospin~\cite{Blaizot1986,Schiro2010,Sciolla2011,Bunemann2013,Seibold2021}.
% One can show that each elementary oscillator has a natural frequency that is parametrically modulated in time with twice the driving frequency as required by point (ii) above (see Appendix~\ref{app:ap3}).

For classical systems it has being proposed to use parametric oscillators as a building block of a time-crystal~\cite{Heugel2019,Yao2020}. In contrast, our building blocks, non-interacting pseudospins, do not
show parametric resonances. The resonances and the time-crystal phase \emph{emerge}  as a result of the interactions between quasiparticles.

Heating and decoherence are often a concern for observing subtle out-of-equilibrium effects in condensates. As shown in Fig.~S3 of Ref.~\cite{Note1}
%Figure~\ref{fig:fig5} shows that
Arnold tongues are visible even if we restrict to relatively few decades of $t\Delta_0$. 
This can be compared with quasiparticle relaxation times $\tau$ of the order of microseconds  measured in aluminium based solid state superconducting devices~\cite{Saira2012} ($\tau\Delta_0\approx 10^6$). While the out-of-equilibrium dynamics may strongly affect this coherence time, the gentle perturbation represented by a subgap drive suggests that the present effects could be observable in solid state superconductors.

The phase diagram is also robust respect to different driving methods. Indeed, we find similar results with a periodic drive of the density of states or driving with an external pairing field which will be shown elsewhere~\footnote{H. P. Ojeda Collado, G. Usaj, C. A. Balseiro, D. H. Zanette, and J. Lorenzana,''Dynamical Phase Transitions in Driven BCS systems'', in preparation.}. The former driving 
can be implemented with ultra cold-atoms, cavity QED or THz radiation in condensed-matter systems with suitable polarizations~\cite{OjedaCollado2018,OjedaCollado2020,Matsunaga2013,Matsunaga2014,Cea2015,Homann2020}.
%In the last case, the coupling arises from the diamagnetic term which is quadratic in the THz field, therefore the drive frequency in the present work, corresponds to twice~\cite{OjedaCollado2018,OjedaCollado2020} the frequency of THz radiation $\omega_d=2\omega_{THz}$. In this way, the $n=1$ resonance found here, corresponds to the familiar   $\omega_d=2\omega_{THz}=2\Delta_0$ found in other works~\cite{Matsunaga2013,Matsunaga2014,Cea2015,Homann2020}. To the best of our knowledge, the resonances for $n>1$ have not being discussed before in the contest of BCS systems.

Periodic $\lambda$-driving can be naturally implemented in ultra-cold atomic gases by using a small time-dependent magnetic field modulation in a Feshbach resonance. 
An alternative protocol has been implemented in cold fermionic lithium atoms~\cite{Behrle2018}. % involving  a radiofrequency field that couples one of the two hyperfine states that simulate spin with a third one.
Parametric resonances, time-crystal phases, and DPTs can be detected through magnetic sweep to the BEC side, giving access to the BCS condensed fraction and the instantaneous order parameter~\cite{Behrle2018}.

Very recently, an optical-cavity QED platform to simulate BCS system has been proposed~\cite{Lewis-Swan2021}. %in which the spin degrees of freedom of trapped atoms play the role of Copper pairs as in the pseudospin Anderson description we have used above~
Such setting is also promising to observe our predictions since it allows for a significant control of Hamiltonian parameters with long coherence times, as already demonstrated in related experiments~\cite{Norcia2018,Muniz2020}.

% %%%%%%%%%%%%%%%%%%%%%%%%%%%%%%%%%%%%%%%%%%%%%%%%%%%%%
% \begin{figure}[tb]
% \includegraphics[width=0.45\textwidth]{Fig5.pdf}
% \caption{(Color online) Short time dynamical phase diagram. We apply  the drive for varying periods of time to show that the main features can be seen before other relaxation process not taken into account could take place. The time window is as follows (a) $t\Delta_0 \in [0, 30]$, (b) $t\Delta_0 \in[0, 60]$, (c) $t\Delta_0 \in[0, 90]$, (d) $t\Delta_0 \in[0, 120]$, (e) $t\Delta_0 \in[0, 150]$ and (f) $t\Delta_0 \in [0, 200]$
% that corresponds with Fig.~\ref{fig:fig1} (a). A convergence of the phase diagram
% at the present resolution can be observed by contrasting panel (e) with (f). 
% On the other hand, increasing even more the time, finer details appear which need however more resolution in the figure to be seen. 
% } 
% \label{fig:fig5}
% \end{figure}
% %%%%%%%%%%%%%%%%%%%%%%%%%%%%%%%%%%%%%%%%%%%%%%%%%%%%%%%%

%\section{Discussion and Outlook}

%One might wonder if the integrability of the model is crucial for the present results. Previous studies~\cite{Scaramazza2019} analyzing the phase diagram of the BCS model for the quench protocol, suggest that this is not the case, but what is crucial is the mean-field nature of the dynamics, which we expect to be generically valid for a large class of physical systems for which weak-coupling theory applies.

The BCS formalism, originally developed for superconductivity,  also applies to  weak-coupling charge- and spin-density waves. Therefore,  our results are also relevant to these types of order  at times short enough for the energy relaxation process to be neglected.

To conclude, we have found that the dynamical phase diagram of a periodically driven BCS condensate is surprisingly rich: it shows Arnold tongues corresponding to parametric resonances mimicking the behavior of a vertically excited pendulum. The dynamics is highly non-trivial showing commensurate (incommensurate) time-translational symmetry breaking inside (outside) the Arnold tongues. This calls for an experimental exploration of the phase diagram. Furthermore, our findings 
%Our contribution introduces a new dimension to the manipulation of BCS condensates by Floquet engineering and  
%represents an incentive to experimentally explore the richness of the phase diagram. The the found nonlinearities
suggest exploring potential applications in parametric amplification, frequency converters and sensing. 

%%%%%%%%%%%%%%%%%%%%%%%%%%%%%%%%%%%%%%%%%%%%%%
\begin{acknowledgments}
  We acknowledge financial support from Italian MAECI and Argentinian MINCYT through  bilateral project AR17MO7, from ANPCyT (grants PICT 2016-0791 and PICT 2018-1509), CONICET (grant PIP 11220150100506), from SeCyT-UNCuyo (grant 06/C603), from Italian Ministry for University and Research through PRIN Project No. 2017Z8TS5B and from Regione Lazio (L.R. 13/08) under project SIMAP. 
HPOC is supported by the Marie Sk\l{}odowska-Curie individual fellowship Grant agreement SUPERDYN No. 893743.
\end{acknowledgments}

% \bibliography{library,library_HP}


\end{document}